\documentstyle[12pt]{article}
\catcode`\@=11

\@addtoreset{equation}{section}
\def\theequation{\arabic{section}.\arabic{equation}}

\def\@normalsize{\@setsize\normalsize{15pt}\xiipt\@xiipt
\abovedisplayskip 14pt plus3pt minus3pt%
\belowdisplayskip \abovedisplayskip
\abovedisplayshortskip  \z@ plus3pt%
\belowdisplayshortskip  7pt plus3.5pt minus0pt}

\def\small{\@setsize\small{13.6pt}\xipt\@xipt
\abovedisplayskip 13pt plus3pt minus3pt%
\belowdisplayskip \abovedisplayskip
\abovedisplayshortskip  \z@ plus3pt%
\belowdisplayshortskip  7pt plus3.5pt minus0pt
\def\@listi{\parsep 4.5pt plus 2pt minus 1pt
            \itemsep \parsep
            \topsep 9pt plus 3pt minus 3pt}}

\def\underline#1{\relax\ifmmode\@@underline#1\else
        $\@@underline{\hbox{#1}}$\relax\fi}
\@twosidetrue
\relax

\catcode`@=12

\catcode`\@=11

\def\section{\@startsection{section}{1}{\z@}{3.5ex plus 1ex minus
   .2ex}{2.3ex plus .2ex}{\large\bf}}

\def\thesection{\Roman{section}.}

\def\appendix{\setcounter{section}{0}
        \def\thesection{APPENDIX }
        \def\theequation{\Alph{section}.\arabic{equation}}}
\def\ps@headings{\def\@oddfoot{}\def\@evenfoot{}
\def\@oddhead{\hbox{}\hfill
        \makebox[.5\textwidth]{\raggedright\ignorespaces --\thepage{}--
        \hfill {\rm }}}
\def\@evenhead{\@oddhead}
\def\subsectionmark##1{\markboth{##1}{}}
}

\ps@headings

\catcode`\@=12

\relax

\def\figcap{\section*{Figure Captions\markboth
        {FIGURECAPTIONS}{FIGURECAPTIONS}}\list
        {Fig. \arabic{enumi}:\hfill}{\settowidth\labelwidth{Fig. 999:}
        \leftmargin\labelwidth
        \advance\leftmargin\labelsep\usecounter{enumi}}}
 \relax
\def\tablecap{\section*{Table Captions\markboth
        {TABLECAPTIONS}{TABLECAPTIONS}}\list
        {Table \arabic{enumi}:\hfill}{\settowidth\labelwidth{Table
 999:}
        \leftmargin\labelwidth
        \advance\leftmargin\labelsep\usecounter{enumi}}}
 \relax
\def\reflist{\section*{References\markboth
        {REFLIST}{REFLIST}}\list
        {[\arabic{enumi}]\hfill}{\settowidth\labelwidth{[999]}
        \leftmargin\labelwidth
        \advance\leftmargin\labelsep\usecounter{enumi}}}
 \relax

\catcode`\@=11

\def\ps@headings{\def\@oddfoot{}\def\@evenfoot{}
\def\@oddhead{\hbox{}\hfill
        \makebox[.5\textwidth]{\raggedright\ignorespaces --\thepage{}--
        \hfill }}
\def\@evenhead{\@oddhead}
\def\subsectionmark##1{\markboth{##1}{}}
}

\ps@headings

\relax
\newskip\humongous \humongous=0pt plus 1000pt minus 1000pt
\def\caja{\mathsurround=0pt}
\def\eqalign#1{\,\vcenter{\openup1\jot \caja
        \ialign{\strut \hfil$\displaystyle{##}$&$
        \displaystyle{{}##}$\hfil\crcr#1\crcr}}\,}
\newif\ifdtup

\def\beq{\begin{equation}}
\def\eeq{\end{equation}}

\def\beqn{\begin{eqnarray}}
\def\eeqn{\end{eqnarray}}
\relax

\def\G2{{\; \rm GeV/}c^2}
\def\G{\; \rm GeV}

\def\dotx{\dotx{\dot\overline{x}}}

\relax

\begin{document}
\hbadness=10000
\begin{titlepage}
\nopagebreak
\begin{flushright}

        {\normalsize KUCP-56\\

        February,~1993}\\
\end{flushright}
\vfill
\begin{center}
{\large \bf q-Deformed Superconformal Algebra \\
on Quantum Superspace}
\vfill
\renewcommand{\thefootnote}{\fnsymbol{footnote}}
{\bf Tatsuo Kobayashi\footnote{
Fellow of the Japan Society for the Promotion of Science. Work
partially supported by the Grant-in-Aid for Scientific Research from
 the
Ministry of Education, Science and Culture (\# 030083)} and
Tsuneo Uematsu\footnote{Work partially supported by the Grant-in-Aid
 for

Scientific Research from the Ministry of Education, Science and Culture

(\# 04245221)}}

       Department of Fundamental Sciences, FIHS, \\
       Kyoto University,~Kyoto 606,~Japan \\
\vfill

\end{center}

\vfill
\nopagebreak
\begin{abstract}
A quantum deformation of 4-dimensional superconformal algebra realized
 on
quantum superspace is investigated.  We study the differential calculus
 and
the action of the quantum generators corresponding to $sl_q(1|4)$ which

act on the quantum superspace.
We derive deformed $su(1|2,2)$ algebras from the deformed $sl(1|4)$
algebra. Through a contraction procedure we obtain a deformed
super-Poincar{\'e} algebra.

\end{abstract}

\vfill
\end{titlepage}
\pagestyle{plain}
\newpage
\voffset = -2.5 cm

Recently there has been a great deal of interest in quantum groups and
quantum algebras \cite{D,J,FRT,KR}
in various fields of theoretical physics and mathematics, such as
 conformal
field theory, integral models, statistical mechanics, knot theory and
 so
on \cite{DH,CFZ,Takhtajan,K}.

Quantum groups can be realized on a quantum space or a quantum
 (hyper-)plane
in which coordinates are non-commuting \cite{Takhtajan,Manin}.
The differential calculus on the non-commutative space
 \cite{Woronowicz}
has been developed by Wess and Zumino \cite{Wess-Zumino} and by other
 people
\cite{Schirrmacher-Wess-Zumino,Schirrmacher1,Schirrmacher2},
especially for $GL_q(n)$
\cite{Schirrmacher-Wess-Zumino,Schirrmacher1,Schirrmacher2},
as well as for $SO_q(n)$ \cite{Carow-Watamura-Schlieker-Watamura}.

So far most of the quantum deformations have been applied to the Lie
 groups
or algebras corresponding to the internal symmetries.
Now it would be interesting to explore quantum deformation of
 space-time
symmetries, which might be relevant at very short distances e.g. at the
Planck length implied by unification theories including gravity.
In the last few years, there have been pioneering works in extending
quantum-group ideas to space-time symmetries including Lorentz,
 Poincar{\'e}
and conformal algebras [16--23].

In our previous paper
\cite{KU}, we studied a quantum deformation of 4-dimensional
conformal algebras based on the quantum space realizing $sl_q(4)$.
Through a contraction procedure we also derived a quantum deformation
 of
Poincar{\'e} algebra.
In the present paper, we shall investigate a q-deformed superconformal
algebra in 4-dimensions in the framework of the quantum superspace
 which

realizes q-deformed superalgebra, $sl_q(1|4)$. By taking a contraction
limit we obtain a deformed super-Poincar{\'e} algebra.

Following \cite{Wess-Zumino,KU} we now consider actions of deformed
$sl(1|4)$ generators on a quantum superspace, whose coordinates $Z^I$
 and derivatives $\partial_I$ consist of bosonic elements $x^0,
 \partial_0$ and fermionic ones $\theta^\alpha, \partial_\alpha$
 $(\alpha=1\sim 4)$.
In Ref.\cite{KU},
the deformed $sl(4)$ algebra has been represented on a 4-dimensional
 quantum bosonic space.
In this paper the same deformed algebra shall be obtained on the
 fermionic space to supersymmetrize the algebra.

Differential calculus on general quantum superspaces with
 multi-parameters has been studied in \cite{KU2}.
Here we set up the following commutation relations of the coordinates
 and the derivaties with one-parameter,
\hskip -.3cm
\renewcommand{\thefootnote}{\fnsymbol{footnote}}
\footnote{Matrices which transform covariantly these commutation
 relations have a superdeterminant as a center (see in detail
 \cite{KU2}).}
$$ Z^IZ^J={(-1)^{\hat I \hat J} \over q}Z^J Z^I, \quad
 \partial_I \partial_J=(-1)^{\hat I \hat J} q \partial_J
 \partial_I,\quad (I<J),$$
$$\partial_I Z^J={(-1)^{\hat I \hat J} \over q} Z^J \partial_I, \qquad
 (I\neq J), \eqno(1)$$
$$\partial_I Z^I=1-(-q)^{2\hat I-2}Z^I \partial_I
 +(q^2-1)\sum_{I<J}Z^J\partial_J,$$
where $\hat I$ denotes the grassmann parity, i.e., $\hat I=0$ for $I=0$
 and $\hat I=1$ for $I \neq 0$.
These commutation relations are written in terms of the following $\hat
 R$-matrix,
$$ \hat R^{IJ}_{\ \ KL}=\delta^I_L \delta^J_K \{
 \delta^{IJ}((-q)^{2\hat
 I}-(-1)^{\hat I \hat J}q)+(-1)^{\hat I \hat J}q) \}+\delta^I_K
 \delta^J_L \Theta^{JI}(1-q^2),
\eqno(2)$$
where $\Theta^{IJ}=1$ for $I>J$ otherwise it vanishes.
For example, we can use the $\hat R$-matrix to write the commutation
 relations between $Z^I$ and $\partial_K$ as
$$ \partial_K Z^I=\delta^I_K+q^{-2}{\hat R}^{IJ}_{\ \ KL}Z^L\partial_J.
\eqno(3)$$
Ordering of the bosonic and fermionic coordinates is non-trivial due to
 $\Theta$.
We study here ordering where $\theta^\alpha$ follows $x^0$, i.e.,
 $\Theta^{0 \alpha}=0$ and $\Theta^{\alpha 0}=1$.
We could discuss the other ordering.

Now we consider actions of the deformed $sl(1|4)$ generators on the
 above quantum space.
First of all, we discuss actions of $T^I_{\ J}$ ($J=I+1,I-1$) which
 correspond to $Z^I \partial_J$ in the \lq \lq classical" limit ($q
 \rightarrow 1$).
Following \cite{Wess-Zumino}, we choose the following basis of the
 algebra,
$$[T^I_{\ J},Z^K]_{a(I,J,K)}=\delta^K_JZ^I, \quad
 [T^I_{\ J},\partial_K]_{b(I,J,K)}=c(I,J)\delta^I_K\partial_J,
\eqno(4)$$
where $[A,B]_a \equiv AB-aBA$ and it is called a q-commutator
 hereafter.
Similarly, a q-anticommutator is defined as $\{A,B\}_a \equiv AB+aBA$.
We investigate consistency between (1) and (4), so as to obtain the
 following solution,
$$ a(I,J,K)=b(I,J,K)^{-1}=- c(I,I+1)^{-1}=q^{2 \hat I-1}, \quad
 (K=min(I,J))$$
$$ a(I,J,K)^{-1}=b(I,J,K)=c(I+1,I)=(-1)^{\hat I+1}q, \quad
 (K=max(I,J))$$
$$ a(I,J,K)=b(I,J,K)=(-1)^{\hat K(\hat I+\hat J)}, \quad (K \neq I,J),
\eqno(5)$$
where $J=I\pm 1$ and $min(I,J)$ ($max(I,J)$) denotes the smaller
 (larger) number in the parenthesis.

Next we define Cartan elements $H_I$ ($I=0\sim 3$) using q-commutators
 of $T^I_{\ I+1}$ and $T^{I+1}_{\ I}$.
We choose $H_I$ as follows,
$$H_0=\{T^1_{\ 0},T^0_{\ 1} \}, \qquad
 H_\alpha=q^{-1}[T^{\alpha+1}_{\ \alpha},T^\alpha_{\ \alpha+1} ]_{q^2},
\eqno(6)$$
for the purpose that terms depending on $T^I_K$ disappear in the
 actions
 of $H_I$.
Note that in the classical limit $H_I=T^{I+1}_{\ I+1}+(-1)^{\hat
 I}T^{I}_{\ I}=Z^{I+1}\partial_{I+1}+(-1)^{\hat I}Z^{I}\partial_{I}$.
It is remarkable that the definition of $H_0$ is not deformed.
The Cartan elements act on the quantum space as follows,
$$[H_0,Z^I]_{1/q^2}=Z^I,\quad
 [H_0,\partial_I]_{q^2}=-q^2\partial_I,\quad (I=0,1),$$
$$[H_\alpha.\theta^\alpha]_{q^2}=-q\theta, \quad
 [H_\alpha,\theta^{\alpha+1}]_{1/q^2}=q^{-1}\theta^{\alpha+1},
\eqno(7)$$
$$[H_\alpha, \partial_\alpha]_{1/q^2}=q^{-1}\partial_\alpha, \quad
 [H_\alpha,\partial_{\alpha+1}]_{q^2}=-q\partial_{\alpha+1},$$
and $H_I$ commute with the other coordinates and the other derivatives.

The other elements of the deformed $sl(1|4)$ can be defined in terms of

q- commutators of $T^I_{\ I+1}$ and
$T^{J+1}_{\ J}$ so that their commutation relations close.
The requirement of the algebraic closure is satisfied if we choose the
 other elements as follows,
$$ T^I_{\ K}=[T^I_{\ J},T^J_{\ K}]_q,\qquad (I<J<K {\rm \ or \ }I>J>K).
\eqno(8)$$
The general non-Cartan generators act on the quantum space as follows,
$$ T^I_{\ K}Z^K=-(-1)^{\hat I}q^{-1}Z^K T^I_{\ K}+Z^I+(-1)^{\hat
 I}(q-q^{-1})\sum Z^JT^I_{\ J},$$
$$T^K_{\ I}Z^I=q^{2\hat I-1}Z^IT^K_{\ I}+Z^K,\quad
T^K_{\ I}Z^J=-(-1)^{\hat I}(Z^JT^K_{\ I}+(q-q^{-1})Z^KT^J_{\ I})$$
$$[T^0_{\ \alpha},x^0]_{1/q}=\{T^\alpha_{\ 0},\theta^\alpha \}_{1/q}=
[T^\alpha_{\ \beta},\theta^\alpha]_{q} =
 [T^\beta_{\ \alpha},\theta^\beta]_{1/q} = 0,\eqno(9)$$
$$T^I_{\ K}\partial_I=q^{1-2\hat I}(\partial_I T^I_{\ K}-\partial_K),
 \quad
T^I_{\ K}\partial_J=-(-1)^{\hat I}\partial_J
 T^I_{\ K}+(q-q^{-1})\partial_KT^I_{\ J},$$
$$ T^K_{\ I}\partial_K=-(-1)^{\hat I}\{ \partial_K T^K_{\ I}
 +q^{2(K-I)-1}\partial_I-(q-q^{-1})\sum q^{2(K-J)}\partial_J
 T^J_{\ I}\},$$
$$ \{T^0_{\ \alpha},\partial_\alpha \}_q=[T^\alpha_{\ 0},
 \partial_0]_q=
[T^\alpha_{\ \beta},
 \partial_\beta]_q=[T^\beta_{\ \alpha},\partial_\alpha]_{1/q}=0,$$
where $I<J<K$ and $\alpha <\beta$, and the non-Cartan generators and
 the
 other elements of the quantum space satisfy the \lq \lq classical"
 algebra, i.e., they commute or anti-commute each other, depending upon
 their grassmann parity.
Using (7) and (9), we can derive commutation relations between the
 deformed $sl(1|4)$ generators.
Some commutation relations among the whole deformed $sl(1|4)$ algebra
 are shown in the following;
$$ [T^I_{\ I+1},T^{I+2}_{\ I+1}]_q = [T^{I+1}_{\ I+2},T^{I+1}_{\ I}]_q
 =
 [T^I_{\ J},T^K_{\ L}]=0, \quad (I,J<K,L),$$
$$[H_0,T^1_{\ \alpha}]_{1/q^2}=T^1_{\ \alpha}, \quad
 [H_0,T^\alpha_{\ 1}]_{q^2}=-q^2T^\alpha_{\ 1},$$
$$[H_\alpha,T^0_{\ \alpha}]_{1/q^2}=q^{-1}T^1_{\ \alpha}, \quad
 [H_\alpha,T^\alpha_{\ 0}]_{q^2}=-qT^\alpha_{\ 0},
\eqno(10)$$
$$[H_\alpha,T^\alpha_{\ \alpha+1}]_{q^4} = -q^2(q+q^{-1})
 T^\alpha_{\ \alpha+1}, \quad
 [H_\alpha,T^{\alpha+1}_{\ \alpha}]_{1/q^4}
 = q^2(q+q^{-1})T^{\alpha+1}_{\ \alpha},$$
$$[H_\alpha,T^\beta_{\ \beta+1}]_{1/q^2}=q^{-1}T^\beta_{\ \beta+1},
 \quad
[H_\alpha,T^{\beta+1}_{\ \beta}]_{q^2}=-qT^{\beta+1}_{\ \beta},\quad
 (\alpha=\beta \pm 1),$$
$$[H_0,T^1_0]=[H_0,T^0_{\ 1}]=[H_I,T^J_{\ K}]=0, \quad (J,K \neq
 I,I+1),$$
and the other commutation relations of the deformed $sl(1|4)$ algebra
 can be derived from the above commutation relations.
It is clear that the Cartan elements commute with each other and
 $T^0_{\ \alpha}$ and $T^\alpha_{\ 0}$ are nilpotent.
In this algebra, the generators $H_\alpha$ and $T^\alpha_{\ \beta}$
 construct $sl_q(4)$ algebra as a closed subalgebra of the deformed
 $sl(1|4)$.
It implies that from the algebra we can derive a deformed
 superconformal
 algebra which includes the deformed conformal algebra as a closed
 subalgebra.
The whole commutation relations of the deformed $sl(4)$ algebra have
 been shown explicitly in \cite{KU} and the whole commutation relations
 of the deformed $sl(1|4)$ algebra will be  explictly shown elsewhere.

{}From the present result, we can easily derive the deformed
 superconformal
algebra in 2-dimensions, $sl_q(1|2)$, by restricting ourselves to the
generators $H_0$, $H_1$, ${T^I}_J \ (I,J=0,1,2)$.
We need another space conjugate to $Z^I$ in order to define the
 deformed
 superconformal algebra $su(1|2,2)$ from the deformed $sl(1|4)$
 algebra.
Here we introduce the conjugate space $\overline Z_I=(\overline
 x_0,\overline \theta_\alpha)$.
Using the $\hat R$-matrix, we set up commutation relations of
 $\overline
 Z_I$ as follows;
$$\overline Z_L \overline Z_K =\hat R^{IJ}_{\ \ KL} \overline Z_J
 \overline Z_I, \qquad \overline Z_k Z^I =\hat R^{IJ}_{\ \ KL} Z^L
 \overline Z_J.
\eqno(11)$$
These commutation relations have a center $C=x^0\overline
 x_0+\sum_\alpha \theta^\alpha \overline \theta_\alpha$.
Further, we assume that the deformed $sl(1|4)$ generators have the same
 actions on $\overline Z_I$ as those on $\partial_I$.
For example, the genarator $T^I_{\ K}$ ($I<K$) acts on $\overline Z_I$
 as follows,
$$T^I_{\ K}\overline Z_I=q^{1-2\hat I}(\overline Z_I
 T^I_{\ K}-\overline
 Z_K).
\eqno(12)$$

Now we consider conjugation between $Z^I$ and $\overline Z_I$.
We assume that $Z^I$ and $\overline Z_I$ are related as
$$ \overline x_0=K_0 (x^0)^*,\quad \overline \theta_\alpha=K_\alpha
 \eta^\alpha (\theta^\alpha)^*,\eqno(13)$$
where $*$ denotes complex conjugate and $K_\alpha$ are coefficients
 which become identity in the classical limit and $\eta^\alpha$ is a
 $su(2,2)$ metric, i.e., $\eta^\alpha=(1,1,-1,-1)$.
The conjugation (13) should be consistent with the commutation
 relations, (1) and (11).
We can find two types of the consistent conjugations depending upon a
 value of the deformation parameter $q$.
At first we discuss the case where $q$ is real.
In this case, the conjugation is allowed if $K_I=1$ and the conjugation
 includes reversing of order, i.e., $\overline {ab}=\overline b
 \overline a$.
Then we take the conjugation of (12) and compare it with (9), so that
 we
 obtain conjugation relations of the generators as follows,
$$ \overline {T^I_{\ J}}=\eta^I\eta^J T^J_{\ I}.
\eqno(14)$$
Similarly, we have
$$ \overline {H_I}=H_I.
\eqno(15)$$

Next we discuss the case with $|q|=1$, where the conjugation does not
 include reversing of order, i.e., $\overline {ab}=\overline a
 \overline
 b$.
This conjugation is consistent with (1) and (11) if $K_\alpha=q^\alpha$
 and $K_0=iq$.
Further we follow the procedure similar to (14) to obtain the following
 conjugation relations;
$$\overline {T^I_{\ J}}=-q^{I-J+1}\eta^I \eta^J T^J_{\ I}, \quad
\overline {H_I}=-q^{2\hat I-2}H_I, \quad (I<J)
\eqno(16)$$
where $\eta^0=i/q$.

We now obtain the quantum deformation of 4-dimensinal superconformal
algebra by assigning the $su_q(1|2,2)$ elements to the physical
 superconformal
generators.
Here we shall consider the case where the deformation parameter $q$ is
 real.
The similar analysis can be performed for the case of $|q|=1$.

We first consider the generators of the conformal sector.
The $SO(4,2)$ generators $M_{mn}$ ($m$,$n$
= $0,1,2,3,4,5$) with the metric $g_{mn}= {\rm
 diag}\ (1,-1,-1,-1,-1,1)$
satisfy
$$[M_{mn},M_{kl}] = -ig_{mk}M_{nl}+ig_{nk}M_{ml} +
 ig_{ml}M_{nk}-ig_{nl}M_{mk}.
\eqno(17)$$
This can be realized by the $su(1|2,2)$ elements as
$M_{mn}={T^\alpha}_\beta{({\sigma}_{mn})_\alpha}^\beta$ (See e.g.
 ref.\cite{PVNS}) where we take
($\mu,\nu = 0,1,2,3$; $\alpha,\beta=1,\cdots,4$)
$$\eqalign{
&\sigma_{\mu\nu}=\textstyle{i\over 4}[\gamma_{\mu},\gamma_{\nu}], \quad
\sigma_{4\mu}=\textstyle{i\over 2}\gamma_{\mu}, \cr
&\sigma_{5\mu}=\textstyle{i\over 2}\gamma_5\gamma_{\mu}, \quad
\sigma_{45}=\textstyle{i\over 2}\gamma_5. \cr
}\eqno(18)$$
The conjugation of ${T^\alpha}_\beta$ (14) and (15) for $q$ real,
 leads to the reality of $M_{mn}$:
$$ \overline{M_{mn}}=M_{mn}
\eqno(19)$$
with the suitable representation for gamma-matrices $\gamma_\mu$
($\mu=0,1,2,3$) which satisfy
$$\eta^\alpha \eta^\beta {({\sigma^*}_{mn})_\alpha}^\beta =
 {({\sigma}_{mn})_\beta}^\alpha.
\eqno(20)$$
Now we regard the 4-th and 5-th dimensions as the compact space.
We construct the generators of classical $4D$ conformal group
out of these rotational generators of $SO(4,2)$.
First, the Lorentz rotation operators $M_{\mu\nu}$ ($\mu,\nu=0,1,2,3$)
is given as $M_{\mu\nu}=M_{m=\mu,n=\nu}$ with
$$\eqalign{
&M_{12}=\textstyle{-1\over 2}(H_1+H_3), \quad
M_{23}=\textstyle{-1\over 2}({T^1}_2+{T^2}_1+{T^3}_4+{T^4}_3), \cr
&M_{31}=\textstyle{i\over 2}({T^1}_2-{T^2}_1+{T^3}_4-{T^4}_3), \quad
M_{01}=\textstyle{-1\over 2}({T^1}_4-{T^4}_1+{T^2}_3-{T^3}_2) \cr
&M_{02}=\textstyle{i\over 2}({T^1}_4+{T^4}_1-{T^2}_3-{T^3}_2),\quad
M_{03}=\textstyle{-1\over 2}(-{T^1}_3+{T^3}_1+{T^2}_4-{T^4}_2) \cr
}\eqno(21)$$
while the translation operators $P_{\mu}$,
conformal boost operators $K_{\mu}$ and dilatation operator $D$ are
 given as
$P_{\mu}=M_{4\mu}+M_{5\mu},\quad K_{\mu}=M_{5\mu}-M_{4\mu}, \quad
D=M_{45}$ with
$$\eqalign{
&M_{40}=\textstyle{-i\over 2}({T^1}_3+{T^3}_1+{T^2}_4+{T^4}_2), \quad
M_{41}=\textstyle{-1\over 2}({T^1}_2+{T^2}_1-{T^3}_4-{T^4}_3) \cr
&M_{42}=\textstyle{i\over 2}({T^1}_2-{T^2}_1-{T^3}_4+{T^4}_3), \quad
M_{43}=\textstyle{-1\over 2}(H_1-H_3)\cr
&M_{50}=\textstyle{-1\over 2}(H_1+2H_2+H_3), \quad
M_{51}=\textstyle{i\over 2}({T^2}_3+{T^3}_2+{T^1}_4+{T^4}_1) \cr
&M_{52}=\textstyle{-1\over 2}({T^2}_3-{T^3}_2-{T^1}_4+{T^4}_1), \quad
M_{53}=\textstyle{i\over 2}(-{T^1}_3-{T^3}_1+{T^2}_4+{T^4}_2) \cr
&M_{45}=\textstyle{-1\over 2}({T^1}_3-{T^3}_1+{T^2}_4-{T^4}_2), \cr
}\eqno(22)$$
We assign the generators ${T^0}_\beta$ and ${T^\beta}_0$ to the
 supercharges
$Q_{\alpha}$, $\overline{Q}_{\dot \alpha}$, $S^{\alpha}$,
$\overline{S}^{\dot \alpha}$ as follows:
$$\eqalign{
&Q_1= \textstyle{\sqrt{2}}({T^0}_1-i{T^0}_3), \quad
\overline{Q}_1= \textstyle{\sqrt{2}}({T^1}_0-i{T^3}_0) \cr
&Q_2= \textstyle{\sqrt{2}}(-{T^0}_2+i{T^0}_4), \quad
\overline{Q}_2= \textstyle{\sqrt{2}}(-{T^2}_0+i{T^4}_0) \cr
&\overline{S}^1= \textstyle{\sqrt{2}}({T^0}_1+i{T^0}_3), \quad
S^1= \textstyle{\sqrt{2}}({T^1}_0+i{T^3}_0) \cr
&\overline{S}^2= \textstyle{\sqrt{2}}(-{T^0}_2-i{T^0}_4), \quad
S^2= \textstyle{\sqrt{2}}(-{T^2}_0-i{T^4}_0) \cr
}\eqno(23)$$
Note that here we have the conjuagation (14).
In 4-dimensional N=1 superconformal algebra, in addition to
 $M_{\mu\nu}$,
$P_{\mu}$, $K_{\mu}$ and $D$ we have another bosonic generator, the
 U(1)

charge, which is defined as
$$A=-\textstyle{1\over 4}(4H_0+3H_1+2H_2+H_3)
\eqno(24)$$

Now we take the same assignment of the generators for the q-deformed
case.  The q-deformed superconformal algebra can be read off from the
commutation relations among the $sl_q(1|4)$ generators.

The classical anticommutation relation of $Q_{\alpha}$ and
$\overline{Q}_{\dot \beta}$ given by
$$
\{ Q_{\alpha}, \overline{Q}_{\dot \beta} \}=
2(\sigma^{\mu})_{\alpha{\dot \beta}}P_{\mu}
\eqno(25)$$
where $(\sigma^{\mu})=({\bf 1},\sigma^i)$ and
$({\bar\sigma}^{\mu})=({\bf 1},-\sigma^i)$,
is deformed as, for example for $\alpha={\dot \beta}=1$,
$$\eqalign{
\{ Q_1, &\overline{Q}_1 \}=
2P_{+}+2[(1-q^4)H_0+(1-q^3)H_1+(1-q)H_2 +i(1-q){T^1}_3 \cr
&+i(1-q^3){T^3}_1
+i\lambda {T^1}_3H_0-i(1-q){T^1}_0{T^0}_3+iq^2\lambda{T^3}_1H_0
+q^3\lambda{T^0}_1{T^1}_0 \cr
&+q\lambda{T^0}_2{T^2}_0+q\lambda H_0H_1+
\lambda H_2H_0
+q^3\lambda{T^2}_3{T^3}_2+q^3\lambda{T^1}_3{T^3}_1
+q\lambda H_1 H_2 \cr
&+q^4\lambda^2{T^1}_2{T^2}_1
-q^3\lambda^2{T^2}_3{T^3}_2H_1
-q^2\lambda^2{T^1}_3{T^3}_1H_0-q^2\lambda^2{T^2}_3{T^3}_2H_0 \cr
&-\lambda^2H_1H_2H_0+q^2\lambda^3{T^2}_3{T^3}_2H_1H_0
-q^3\lambda^3{T^1}_2{T^2}_1H_0 ]\cr
}\eqno(26)$$
where $P_{\pm}=P_0\pm P_3$ and $\lambda=q-q^{-1}$.
The classical relation $\{ {\bar S}^{\dot \alpha},S^{\beta} \}
=2({\bar\sigma}^{\mu}K_{\mu})^{{\dot \alpha}\beta}$ is similarly
deformed. While another classical anticommutation relation
$$\{ Q_{\alpha},S^{\beta} \} =
{(\sigma^{\mu\nu})_{\alpha}}^{\beta}M_{\mu\nu} +
 2i{\delta_{\alpha}}^{\beta}D
-4{\delta_{\alpha}}^{\beta}A
\eqno(27)$$
receives q-deformation, e.g. for $\alpha=\beta=1$, as given by
$$\eqalign{
\{ Q_1, &S^1 \}=
{(\sigma^{\mu\nu})_1}^1 M_{\mu\nu}+2iD-4A
+2[(q^4-1)H_0+(q^3-1)H_1+(q-1)H_2  \cr
&+i(1-q){T^1}_3-i(1-q^3){T^3}_1
+i\lambda {T^1}_3H_0-i(1-q){T^1}_0{T^0}_3-iq^2\lambda{T^3}_1H_0  \cr
&-q^3\lambda{T^0}_1{T^1}_0-q\lambda{T^0}_2{T^2}_0-q\lambda H_0H_1-
\lambda H_2H_0 -q^3\lambda{T^2}_3{T^3}_2-q^3\lambda{T^1}_3{T^3}_1 \cr
&-q\lambda H_1 H_2-q^4\lambda^2{T^1}_2{T^2}_1
+q^3\lambda^2{T^2}_3{T^3}_2H_1
+q^2\lambda^2{T^1}_3{T^3}_1H_0+q^2\lambda^2{T^2}_3{T^3}_2H_0 \cr
&+\lambda^2H_1H_2H_0-q^2\lambda^3{T^2}_3{T^3}_2H_1H_0
+q^3\lambda^3{T^1}_2{T^2}_1H_0 ]\cr
}\eqno(28)$$
where $\sigma^{\mu\nu}=\textstyle{i\over
 2}(\sigma^{\mu}{\bar\sigma}^{\nu}-
\sigma^{\nu}{\bar\sigma}^{\mu})$ and
${\bar\sigma}^{\mu\nu}=\textstyle{i\over
 2}({\bar\sigma}^{\mu}\sigma^{\nu}-
{\bar\sigma}^{\nu}\sigma^{\mu})$.

The anti-commutator $\{ Q_{\alpha},Q_{\beta}\}$ is also deformed. We
 find
$$
\{Q_1,Q_2\}_q=2(1-q^2){T^0}_3{T^0}_2, \quad
Q_1^2=2i(q-1){T^0}_3{T^0}_1, \quad
Q_2^2=2i(q-1){T^0}_4{T^0}_2,
\eqno(29)$$
where we note that $Q_{\alpha}$ is no more nilpotent.
We further observe that $Q_\alpha^3=0$, which has a common feature with
 parafermion statistics.
It would be very interesting to investigate the representation theory
 of
 this q-deformed algebra.

Since super-Poincar{\'e} algebra is not a simple Lie superalgebra, we
 cannot
apply the standard prescription of quantum deformation.  As in the case
 of
Poincar{\'e} algebra, we need some contraction procedure. Here we shall
 adopt
the contraction method using rescaling of the translation as well as
supersymmetry generators and subsequent limiting procedure \cite{CGST}.

We first make the following rescaling of the generators
$$\eqalign{
&{\hat M}_{\mu\nu} = M_{\mu\nu}, \quad
{\hat P}_{\mu}={1\over R}P_{\mu}, \quad
{\hat K}_{\mu}=K_{\mu}, \quad
{\hat D}=D, \quad
{\hat A}=A, \cr
&{\hat Q}_{\alpha} = {1\over \sqrt{R}}Q_{\alpha}, \quad
{\hat {\bar Q}}_{{\dot \alpha}} = {1\over \sqrt{R}}{\bar Q}_{{\dot
 \alpha}},
\quad
{\hat S}^{\alpha} = S^{\alpha}, \quad
{\hat {\bar S}}^{{\dot \alpha}} = {\bar S}^{{\dot \alpha}}. \cr
}\eqno(30)$$
Let us substitute the above rescaled generators into the q-deformed
superconformal algebra and then take the contraction limit:
$$
R \rightarrow \infty , \qquad q \rightarrow 1,
\eqno(31)$$
with $\kappa =(q-q^{-1})R^2$ kept fixed.  The $\kappa$ is a new
 deformation
parameter similar to those introduced in refs.
\cite
{Lukierski-Ruegg-Nowicki-Tolstoy,LN,Lukierski-Nowicki-Ruegg,CGST,KU}.

We then get a quantum deformation of super-Poincar{\'e} algbera. In the
following we omit the hat again.
In the sector of the bosonic generators, only the commutation relations
among the Lorentz generators, $[M_{\mu\nu},M_{\lambda\sigma}]$,
receive the deformation which was discussed in
the previous paper \cite{KU}.
The other commutation relations, e.g., $[M_{\mu\nu},P_{\sigma}]$,
 $[P_{\mu},P_{\nu}]$ and $\{ Q_{\alpha}, \overline{Q}_{\dot \beta} \}$
 etc.
remain classical.

In this paper we have investigated the q-deformed $4D$ superconformal
 and
super- Poincar{\'e}
algebras in the framework of the quantum space for $sl_q(1|4)$.  We
 studied
the differential calculus on the non-commutative quantum space for
 $sl_q(1|4)$
as well as the action of the generators on this quantum space.
Through the charge conjugations, we obtain two types of the deformed
$su(1|2,2)$ algebras from the deformed $sl(1|4)$.

We presented the q-deformed superconformal algebra with a suitable
 assignment
of the $4D$ superconformal generators.
The algebra includes the deformed conformal algebra as a closed
 subalgebra.
A quantum super-Poincar{\'e} algebra with a deformation parameter
 $\kappa$
was obtained by rescaling of the translation and supercharge operators
and subsequent contraction procedure.

In the present paper we only discussed the 4-dimensional case
 corresponding
to $sl_q(1|4)$. As we mentioned before, it is quite easy to derive the
2-dimensional case, $sl_q(1|2)$, which is a deformation of the
subalgebra of super Virasoro algebra.
As a future problem, it would be very much interesting to extend the
 present
analysis to the extended superconformal algebras.

\vspace{0.8 cm}
\leftline{\large \bf Acknowledgement}
\vspace{0.8 cm}

The authors would like to thank P.~P.~Kulish, R.~Sasaki and
 C.~Schwiebert
for valuable discussions.

\end{document}